\documentclass{article}
\usepackage{smc2025}
\usepackage[caption=false, font=footnotesize]{subfig}
\usepackage{paralist}
\usepackage[figure,table]{hypcap}
\usepackage{multirow}

\usepackage[whole]{bxcjkjatype}

\usepackage{alphabeta}
\usepackage{arabtex}
\usepackage[LFE,LAE,LGR,T2A,T1]{fontenc}
\usepackage[greek, russian, main=english]{babel}

\def\papertitle{Music interpretation and emotion perception: A computational and neurophysiological investigation}

\author[1]{\mbox{\firstname{Vassilis}\lastname{Lyberatos}\email{vaslyb@ails.ece.ntua.gr}}$^*$}
\author[1]{\mbox{\firstname{Spyridon}\lastname{Kantarelis}\email{spyroskanta@ails.ece.ntua.gr}}$^*$}
\author[2]{\mbox{\firstname{Ioanna}\lastname{Zioga}}\email{joannazioga@gmail.com}}
\author[2]{\mbox{\firstname{Christina}\lastname{Anagnostopoulou}}\email{chrisa@music.uoa.gr}}
\author[1]{\mbox{\firstname{Giorgos}\lastname{Stamou}}\email{gstam@cs.ntua.gr}}
\author[2]{\mbox{\firstname{Anastasia}\lastname{Georgaki}}\email{georgaki@music.uoa.gr}\author[2]}

\affil[1]{\institution{School of Electrical and Computer Engineering, National Technical University of Athens}\city{Athens}\country{Greece}\affiliationtype{University}}
\affil[2]{\institution{Department of Music Studies, National and Kapodistrian University of Athens}\city{Athens}\country{Greece}\affiliationtype{University}}
\affil[*]{These authors contributed equally to this work.}

\completesetup

\title{\papertitle}
\begin{document}
\capstartfalse
\maketitle
\capstarttrue

\begin{abstract}
This study investigates emotional expression and perception in music performance using computational and neurophysiological methods. The influence of different performance settings, such as repertoire, diatonic modal etudes, and improvisation, as well as levels of expressiveness, on performers' emotional communication and listeners' reactions is explored. Professional musicians performed various tasks, and emotional annotations were provided by both performers and the audience. Audio analysis revealed that expressive and improvisational performances exhibited unique acoustic features, while emotion analysis showed stronger emotional responses. Neurophysiological measurements indicated greater relaxation in improvisational performances. This multimodal study highlights the significance of expressivity in enhancing emotional communication and audience engagement.

\end{abstract}

\section{Introduction}\label{sec:introduction}

In recent years, the study of music performance has become a prominent area of research. While traditional analysis of music often relied on the score, modern research highlights the importance of performance-specific features that distinguish one rendition from another. These include subjective decisions made by performers (either consciously or automatically), such as tempo variations, dynamic changes, and articulation ~\cite{lerch2021audio}, which can drastically alter a listener’s perception of the music~\cite{rink2002musical}. Recent findings suggest that the combination of both audio and visual components of a performance significantly enhances the experience and emotional impact~\cite{platz2012eye}.

The goal of audio performance analysis is to identify performance characteristics, often using statistical methods and audio feature extraction techniques. Crucially, the performers’ intentions influence their expressive choices, which are reflected in timing, dynamics, and emotional expression. For example, tempo deviations, such as ritardando or accelerando are typically associated with phrase boundaries~\cite{lerch2019music}. Similarly, dynamic variations, such as loudness fluctuations, are crucial for emphasizing emotional content, as performers manipulate volume to create contrasts and highlight specific sections of a piece. Additionally, pitch-related parameters like vibrato have been found to be particularly important in string and vocal performances, with studies showing that vibrato rate and depth correlate with perceived emotional expression~\cite{trevor2022musicians}.

Expressive performance has been suggested to serve a variety of functions. Its primary role is to convey the performer’s interpretation of the musical structure and emotional content tied with a musical piece. 
Recent studies utilizing electroencephalography (EEG) and physiological measures such as heart rate and skin conductance have provided deeper insights into emotional perception during music performance. For instance, research has demonstrated that EEG-based connectivity in beta and gamma frequency ranges effectively reflects the networks involved in emotional transfer during musical performances. In contrast, low-frequency bands (delta, theta, alpha) were found to be less indicative of these emotional networks~\cite{ghodousi2022eeg}. Another study observed that listening to disliked music evoked physiological reactions indicative of higher arousal, including increased heart rate, skin conductance response, and body temperature, highlighting the complex emotional dynamics elicited by music ~\cite{merrill2023effects}. Another study in 2024 investigated the impact of different musical modes on physiological indicators such as skin temperature, heart rate, and electrodermal activity. The findings revealed that pentatonic mode music led to greater skin temperature changes compared to major and minor modes, suggesting that musical mode influences autonomic nervous system activity, thereby affecting emotional experiences ~\cite{jiang2024inductive}.

Overall, musical communication has been found to involve a complex interaction between the transmitter (the performer), the medium (the instruments and the music itself), and the receiver (the listener). However, despite these advances, there is still much to explore about how both music-related and music-unrelated factors influence the perception of emotion. Our study aims to investigate how these factors contribute to the perception of and emotional responses to music, in order to deepen our understanding of the interplay between emotion expression and perception.

One emergent computational methodology for analyzing complex interactions between features in multimodal settings is machine learning~\cite{baltruvsaitis2018multimodal}. A particularly relevant subfield, eXplainable Artificial Intelligence (XAI), focuses on developing interpretable models that enhance our understanding of both data and model behavior~\cite{guidotti2021principles}. While machine learning has been applied to study emotional perception and musical perceptual features~\cite{lyberatos2024perceptual}, its application to music performance analysis in multimodal settings remains relatively unexplored. In this work, we employ interpretable machine learning techniques to analyze music performance data, aiming to gain deeper insights into the relationship between expressive musical features and performance interpretation. This approach provides a more transparent framework for understanding the expressive nuances of music performance and how performers communicate emotions through their playing.

\subsection{The witheFlow Project}\label{subsec:withtheFlow}

The witheFlow project~\footnote{\url{https://witheflow.ails.ece.ntua.gr/}} presents a novel integration of artificial intelligence (AI) with musical performance, focusing on augmenting human emotional expression through an interdisciplinary approach combining machine learning, music theory, and biometric analysis~\cite{filandrianos2020brainwaves}. The system architecture incorporates real-time biosignal processing algorithms to interpret emotional valence during musical performances while utilizing a comprehensive dataset collected from professional musicians in a controlled acoustic environment. Through a three-phase implementation protocol-encompassing dataset compilation, emotional context mapping, and system development-the project demonstrates significant potential for advancing human-AI collaborative artistic expression. Initial findings suggest promising applications in enhancing musical composition and performance, contributing to the establishment of new paradigms in creative interaction between performers and intelligent systems, while maintaining the primacy of human emotional expression in artistic creation.

\subsection{Study aim and research questions}\label{subsec:aim}
In this study, we aim to investigate emotional expression in music performance and improvisation using behavioural, computational, and neurophysiological measures. We employ a carefully-designed experimental procedure to address the following key questions:

\begin{enumerate}
    \item Which acoustic features  distinguish different types of musical performance?
    \item Which emotions are expressed and perceived in different types of musical performance?
    \item Which acoustic and neural features are associated with emotion expression during music performance?
\end{enumerate}

\section{Methods}\label{sec:methods}

The methods described below detail our approach to investigating music-induced emotions through a multimodal framework combining audio analysis, emotion annotation, and biosignal recording. Our experimental design captured the interaction between performers and listeners in a controlled studio environment, allowing for comprehensive data collection across multiple modalities. The subsequent analysis employed both statistical and machine learning methods to understand the relationships between performance conditions, emotional responses, and physiological signals.

\subsection{Participants}\label{subsec:participants}

Each recording session involves three distinct roles: the research team, the professional musician, and the volunteers. The research team is responsible for data acquisition, training volunteer annotators, and overseeing the experimental procedure. The professional musician, who is compensated for their participation, performs while wearing biosignal recording devices to capture electroencephalography (EEG) and electrocardiography (ECG) signals. Among the volunteers (listeners), one serves as the audience member, passively listening to the  performance while wearing EEG, ECG, and galvanic skin response (GSR) sensors, while the remaining volunteers (typically two or three) annotate the performance in real time (annotator).

The study focused on a subset of 16 sessions. The instruments played by the 16 performers included 2 grand pianos, 2 accordions, 3 saxophones, 1 double bass, 2 electric guitars, 1 classical guitar, 1 electric bass, 1 violin, 1 lute, 1 ney, and 1 Cretan lyra. A total of 34 annotators participated, along with 15 audience members. Fourteen male and two female professional musicians participated in the performance, with a mean age of 52.13 years (SD = 11.61). The performers had an average of 39.76 years (SD = 7.70) of musical experience. Among listeners, there were 30 males and 19 females, with a mean age of 26.56 years (SD = 5.10).

\subsection{Material}\label{subsec:material}
    
\subsubsection{Audio equipment and data description}

All recording sessions take place in a state-of-the-art university laboratory recording studio. High-fidelity microphones are used for acoustic instruments, while dedicated amplifiers capture electric instruments. All performers use their own instruments and equipment (e.g., effect pedals), except for pianists, who perform on the studio's grand piano. The audio is recorded in an uncompressed mono WAV format using 24-bit PCM encoding, with a 48 kHz sampling rate and a bitrate of 1152 kbps, ensuring high-resolution audio quality.

\subsubsection{Annotation equipment and data description} 
 
In each session, two or three volunteers annotate the performance of a professional musician.  The annotation process consists of two components: (1) continuous annotation based on Russell’s circumplex model of affect~\cite{russell1980circumplex}, which represents valence on the x-axis and arousal on the y-axis, and (2) categorical annotation using the nine-factor Geneva Emotional Music Scale (GEMS-9)~\cite{jacobsen2024assessing}, where all participants are instructed to select at least one emotion, with a \textit{Neutral} option also available. Additionally, annotators rate each music segment on a five-point scale to indicate their level of enjoyment and familiarity. Annotations are collected through our custom web-based platform~\footnote{\url{https://witheflow.ails.ece.ntua.gr/annotator/}}. Our study utilizes the categorical annotations. 

The GEMS-9 was employed to assess self-reported emotions felt during music listening. This scale has been demonstrated to comprise a domain-specific tool for measuring emotions induced by music~\cite{zentner2008emotions}. Crucially, unlike perceived musical emotions, which refer to how emotions are interpreted, the GEMS focuses on felt emotional responses~\cite{juslin2004expression}. The scale includes both aesthetic emotions like \textit{Wonder}, and basic emotions like \textit{Sadness}. Finally, the GEMS organizes emotions into hierarchical levels, including second-order factors like \textit{Sublimity}, \textit{Vitality}, and \textit{Unease}.

\subsubsection{Biosignal equipment and data description} 
Biosignals are collected using a wearable EEG device with four dry electrodes. The locations of the electrodes on the scalp correspond approximately to O1, O2, T3, and T4, according to the International 10–20 System. The EEG device also includes one reference electrode and one common sensor, all operating at a 250 Hz sampling rate. Additionally, a separate wearable device records respiration and heart function through ECG and GSR. Our study focuses on the analysis of the EEG signals. 


\subsection{Experimental procedure}\label{subsec:exp_procedure}

The experimental procedure lasts approximately three hours, including one hour for audio setup, annotator training, and biosignal device preparation, followed by two hours of audio recording and live annotation. The session consists of four phases: (1) calibration, where resting-state biosignals are collected; (2) ground repertoire, where all musicians perform the same pieces; (3) personal choice repertoire, featuring two public domain pieces selected by each musician; and (4) improvisation, both free and over a continuous single-note drone, selected by the performer. After each segment, musicians categorize their emotions using GEMS-9 and, at the end of each phase, reflect on the techniques used to enhance expressivity. Performers were asked to name emotion tags based on their perception of the emotions their performance elicited in the audience, while the audience selected tags reflecting their own emotional response to the performance. The three performing phases (2–4) are selected randomly to mitigate biases such as participant fatigue and annotation drift.

The recording sessions are subdivided into discrete tasks, each lasting two to three minutes. The first section includes nine tasks: seven etudes based on the seven diatonic modes (Ionian to Locrian) composed by the research team and two performances of Greensleeves-one in a non-expressive manner and one in an expressive manner. The second section consists of four tasks, where performers record two public-domain pieces of their choice, each performed both non-expressively and expressively. The third section comprises two tasks: a free improvisation and an improvisation over a sustained single-note drone selected by the performer.

\subsection{Data Analysis}\label{subsec:data}

Our study examined over 200 audio tracks alongside EEG signals recorded from both performers and audience members, supplemented by more than 200 categorical annotations from performers and approximately 800 from listeners. The analysis was structured into three distinct settings, each comparing performer and audience perspectives. The Performance Conditions setting explored differences across three performance scenarios-repertoire, modal transformations, and improvisation-through statistical analysis of audio features, explicit emotional annotations, and implicit emotional responses (EEG). The Expressiveness setting focused on variations between expressive and non-expressive performances of repertoire pieces, applying the same analytical approach. Finally, the Holistic setting integrated all collected data, leveraging machine learning techniques to examine complex relationships among audio features, emotional responses, and physiological signals. The following sections outline the specific analytical procedures used in each setting.

\subsubsection{Audio Feature Analysis}
We conducted a comprehensive audio signal analysis to measure and compare variations across different performance conditions. The analysis, performed with a frame size of 4096 samples and hop size of 2048 samples, extracted key features including spectral characteristics (flatness, flux, complexity), rhythmic elements (BPM, attack time, attack slope, pulse clarity), tonal features (pitch salience), and dynamics. Features were analyzed across three frequency bands: low (20-160 Hz), middle (160-1280 Hz), and high (1280-20480 Hz), providing detailed characterization of the musical signal across different octave ranges.

\subsubsection{Emotional Transmission Analysis}
Emotional transmission in music performance represents the process by which performers' emotional intentions are conveyed to and perceived by the audience. This complex phenomenon was investigated through both explicit behavioral measures and implicit neurophysiological responses, allowing us to examine the effectiveness of emotional communication across different performance conditions.

\vspace{1em}

\noindent\textit{Explicit Emotional Analysis}: The explicit component of emotional transmission was assessed through the alignment between performer-intended emotions and audience-perceived emotions. This was achieved by analyzing categorical emotion tags derived from the GEM-9 and the \textit{Neutral} tag, as provided by both performers and audience members during the experimental sessions.

\vspace{1em}

\begin{figure*}[!h]
    \begin{minipage}{0.5\textwidth}
        \centering
        \includegraphics[width=\textwidth]{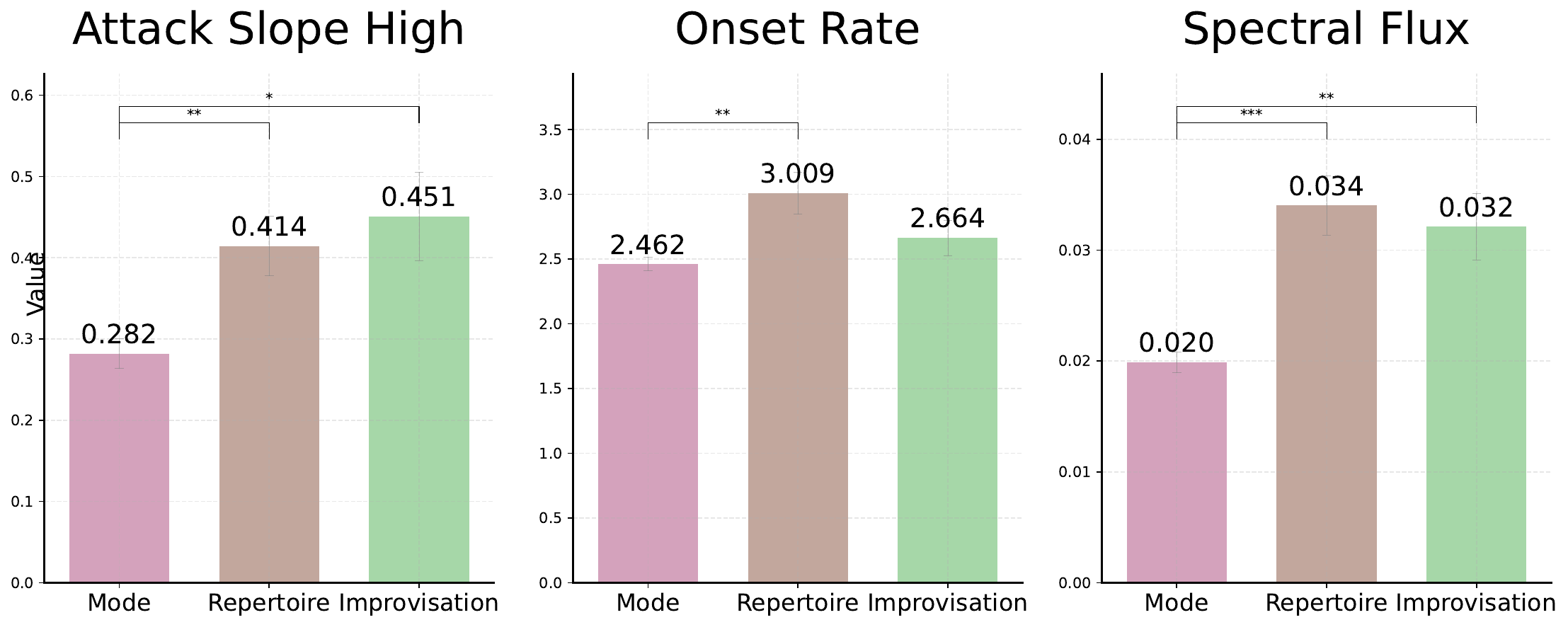}
    \end{minipage}
    \hfill
    \begin{minipage}{0.5\textwidth}
        \centering
        \includegraphics[width=\textwidth]{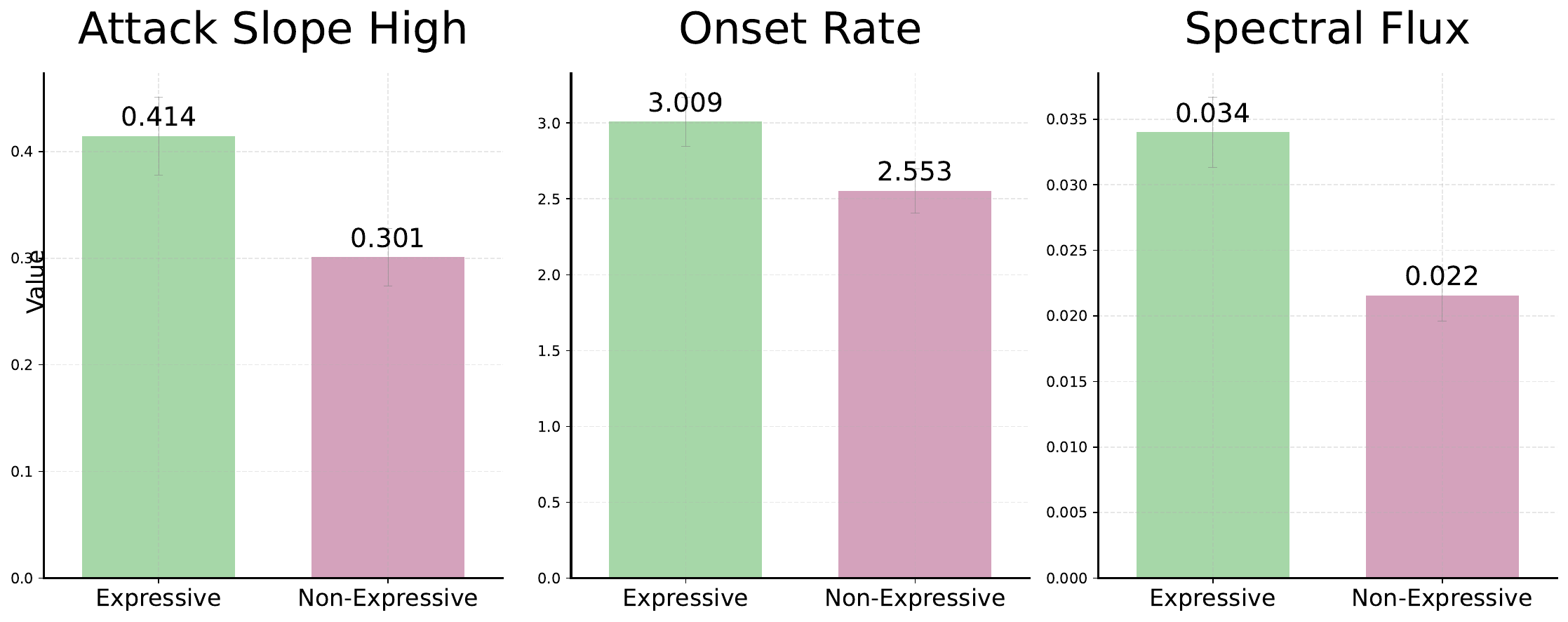}
    \end{minipage}
    \caption{Comparison of statistically significant acoustic features. Left: Analysis across performance conditions (Modes, Repertoire, Improvisation). Right: Analysis between Expressive and Non-Expressive performances.}
    \label{fig:features_comparison}
\end{figure*}

\noindent\textit{Implicit Emotional Analysis}: The implicit component examined neurophysiological alignment using EEG recordings (250 Hz sampling rate) from occipital and temporal channels, specifically focusing on the alpha power shift (8-13 Hz) feature. This measure represents the percentage change in alpha power of the EEG signal compared to the resting state. These objective measurements provided insights into emotional resonance between performers and audience members across different performance conditions and expressive states, complementing the explicit emotional assessments.

\subsubsection{Multimodal Analysis through Machine Learning}

We employed interpretable machine learning techniques, specifically \textsc{Decision Trees}~\cite{kotsiantis2013decision}, to explore the relationships between emotions, audio features, and biosignals. Two distinct datasets were analyzed: one derived from audience responses and one from performer responses. Each dataset was divided into 80\% training and 20\% test sets. Individual Decision Tree models were trained for each emotion category, creating separate models for audience members and performers. The transparent nature of \textsc{Decision Trees} allowed us to visualize which features most strongly influence emotional perception and how these differ between performers and listeners, providing insights into the mechanisms of emotional communication in music.

\section{Results}\label{sec:results}
Our analysis revealed distinct patterns in musical performance and emotional communication across three complementary dimensions. First, we examined acoustic features to understand how different performance conditions and expressiveness levels manifest in the musical signal. Second, we analyzed emotional transmission through both explicit (emotion tags) and implicit (EEG) measures. Finally, we employed machine learning techniques to model the relationships between acoustic features, physiological responses, and emotional perception. Below, we present our findings in detail.
\subsection{Audio Feature Analysis}\label{subsec:audio_features}
We extracted 40 acoustic features using the Essentia~\cite{bogdanov2013essentia} audio analysis library, encompassing various aspects of musical expression including dynamics (loudness, energy), rhythm (BPM, onset rate, pulse clarity), timbre (spectral features), and articulation (attack time and slope). These features were analyzed across frequency bands (low, middle, high).

\subsubsection{Performance Conditions Comparison}
To analyze differences in audio features across performance conditions, we conducted one-way ANOVA followed by post-hoc t-tests. Using Bonferroni correction, the analysis revealed three significantly distinct features. Attack slopes were highest during improvisation, indicating sharper note attacks compared to repertoire and modes. Repertoire performances exhibited the highest onset rates and spectral flux, reflecting more frequent note events and greater timbral variation than both improvisation and modes (Figure \ref{fig:features_comparison}).

\subsubsection{Expressiveness Comparison}

Analysis of expressive versus non-expressive performances using paired t-tests with Bonferroni correction identified significant differences in three acoustic features. Expressive performances demonstrated consistently higher values in attack slope, onset rate, and spectral flux. These findings indicate that expressive playing is characterized by sharper note attacks, more frequent note events, and greater timbral variation compared to non-expressive performances (Figure \ref{fig:features_comparison}).

\begin{table}[!b]
    \centering
    \begin{tabular}{c c c c}
        \hline
        \multirow{2}{*}{\underline{Emotion Tag}} & \multicolumn{3}{c}{\underline{Occurrences (Percentage in \%) }} \\ 
        & \underline{Listeners}  & \underline{Performers}  & \underline{Total} \\ 
        Peacefulness & \textbf{251 (16.2)} & 37 (10.4) & \textbf{288 (15.1)} \\ 
        Sadness & 206 (13.3) & 23 (6.4) & 229 (12.0) \\ 
        Joy & 181 (11.7) & 46 (12.9) & 227 (11.9) \\ 
        Nostalgia & 177 (11.4) & 45 (12.6) & 222 (11.6) \\ 
        Tension & 168 (10.9) & 40 (11.2) & 208 (10.9) \\ 
        Tenderness & 147 (9.5) & 38 (10.6) & 185 (9.7) \\ 
        Power & 143 (9.2) & 34 (9.5) & 177 (9.3) \\ 
        Neutral & 102 (6.6) &\textbf{ 50 (14.0)} & 152 (8.0) \\
        Wonder & 95 (6.1) & 23 (6.4) & 118 (6.2) \\ 
        Transcendence & 78 (5.0) & 21 (5.9) & 99 (5.2) \\ \hline
        Total  & 1548 (100)   & 357 (100) & 1905 (100) \\ \hline
    \end{tabular}
    \caption{Emotion tag frequencies.}
    \label{tab:emotion_occurrences}
\end{table}

\subsection{Emotional Transmission }\label{subsec:emotion_transmission}

We examined emotional communication in music performance through complementary behavioral and neurophysiological measures. This dual approach revealed patterns in how emotions are conveyed and perceived across different performance conditions, providing insights into both conscious emotional responses and underlying physiological processes during musical interaction

\subsubsection{Explicit Emotional Analysis}

A total of 1,905 emotion tags were collected from audience and performers, categorizing their perceived emotional responses to the musical performances. As shown in Table~\ref{tab:emotion_occurrences}, the most frequently assigned tags include \textit{Peacefulness}, \textit{Tension}, and \textit{Joy}, suggesting a prevalence of contrasting affective states. Conversely, emotions such as \textit{Wonder} and \textit{Transcendence} appear less frequently. Notably, \textit{Neutral} is the most frequently selected tag among performers. 

\begin{figure*}[!h]
    \begin{minipage}{0.48\textwidth}
        \centering
        \includegraphics[width=\textwidth]{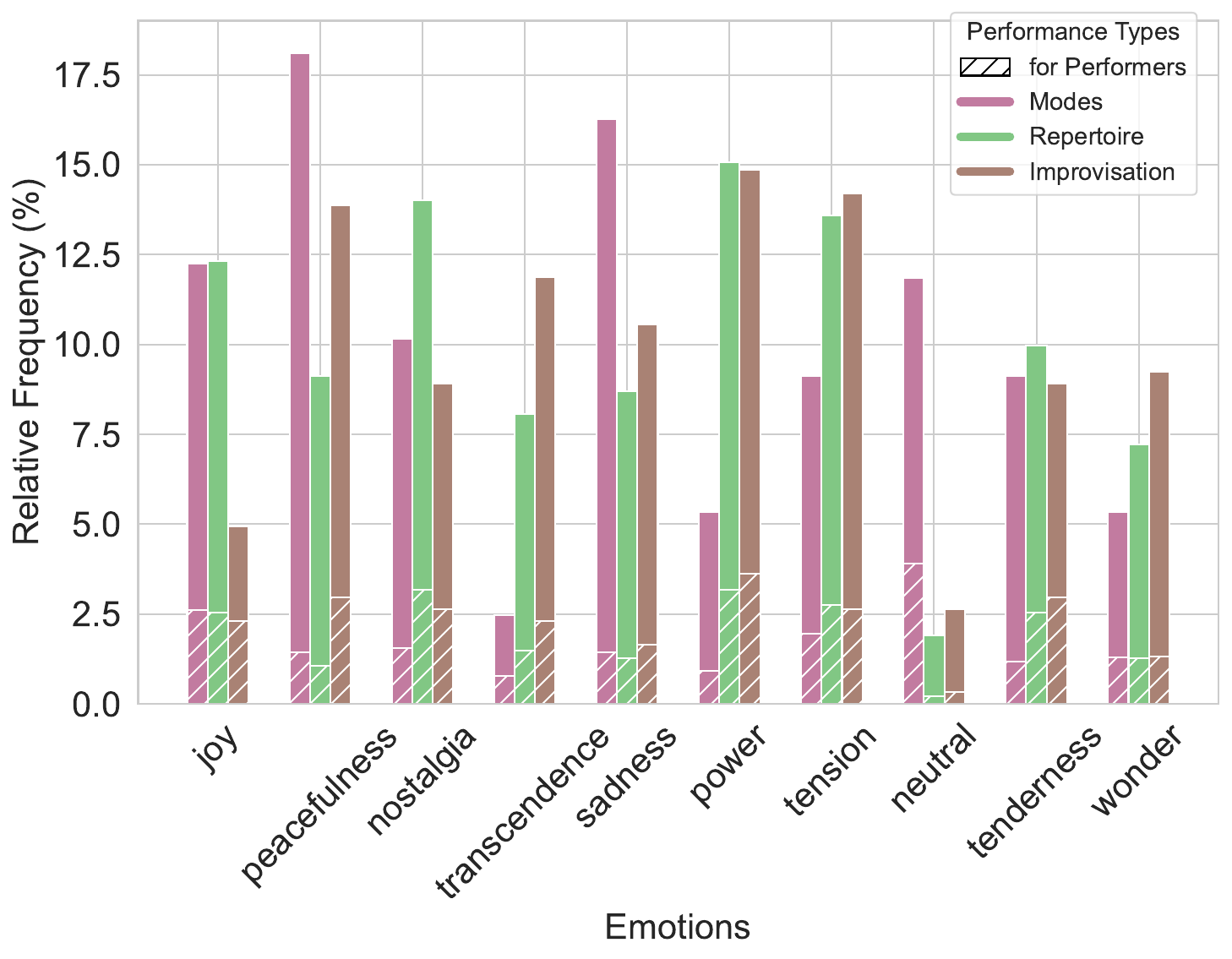}
    \end{minipage}
    \begin{minipage}{0.48\textwidth}
        \centering
        \includegraphics[width=\textwidth]{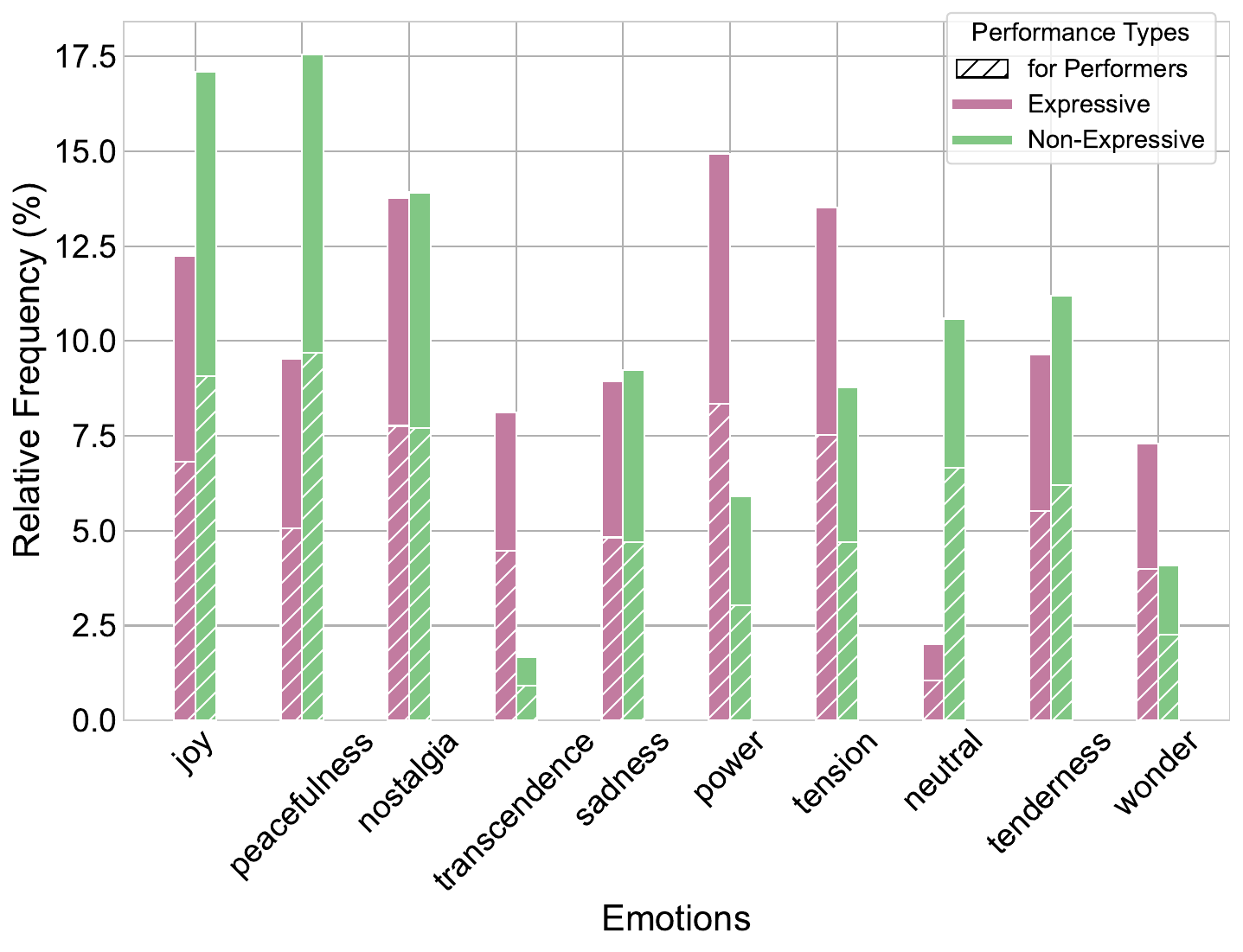}
    \end{minipage}

    \caption{Distribution of reported emotions. Left: Comparison across performance conditions (Modes, Repertoire, Improvisation). Right: Comparison between Expressive and Non-Expressive performances.}
    \label{fig:emotion_frequencies}
\end{figure*}

Figure~\ref{fig:emotion_frequencies} illustrates the frequency of emotion tags in the Expressiveness setting and in the Performance Condition setting, accounting for the varying number of tasks across conditions. The low occurrence of the \textit{Neutral} tag in Expressive and Improvisation tasks, compared to Modes and Non-Expressive tasks, suggests that the former conditions are more effective at evoking emotions. Additionally, high-arousal emotions such as \textit{Power} and \textit{Tension} are more prominent in the Expressive and Improvisation tasks. Furthermore, \textit{Wonder} and \textit{Transcendence}, two less frequently selected tags indicating strong emotional stimuli, are notably present in these conditions as well. 

Lift is a metric that quantifies the strength of association between emotion tags and performance conditions. It is defined as:

\begin{equation*}
\text{Lift}(\text{Emotion} \Rightarrow \text{Expressive}) = \frac{P(\text{Expressive} \mid \text{Emotion})}{P(\text{Expressive})}
\end{equation*}

\begin{figure}[!b]
\centering
\includegraphics[width=1\columnwidth]{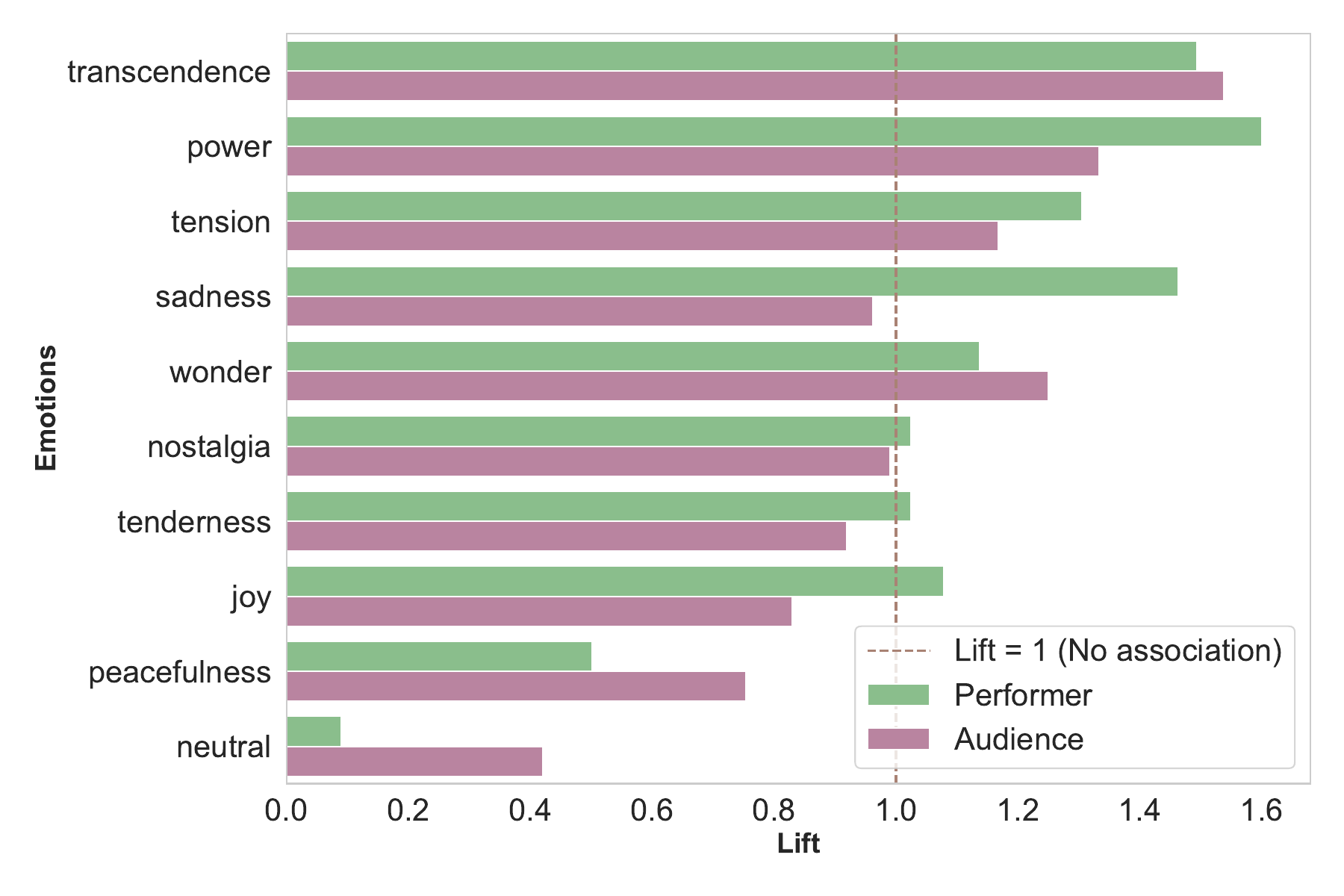}
\caption{Lift of Emotions in the Non-Expressive to Expressive Transition.}
\label{fig:lift}
\end{figure}

A Lift value greater than one indicates that an emotion is more strongly associated with expressive performances than expected, while a value below 1 suggests a weaker association. Figure~\ref{fig:lift} shows that, as performances transition from non-expressive to expressive, the four high-arousal emotions exhibit the greatest Lift, reflecting a strong association with expressive performances. In contrast, \textit{Neutral} has the lowest Lift, indicating a weaker connection to expressive performances, particularly in relation to the emotion tags provided by the audience. Notably, \textit{Sadness} has a Lift value greater than one for performers but remains close to one for the audience, indicating that performers use \textit{Sadness} more to communicate expressivity.

Chi-Square tests were used to assess the relationship between emotional tags assigned by performers and those provided by listeners (audience and annotators) across four conditions: Modes, Improvisation, Expressive Repertoire, and Non-Expressive Repertoire. A high p-value indicates no significant difference between performer intentions and listener perceptions, while a low p-value suggests divergence.

As shown in Table~\ref{tab:chi_square_results}, the Modes condition exhibited the highest disagreement, remaining significant even after applying a Bonferroni correction. Similarly, the Non-Expressive Repertoire condition showed notable divergence. In contrast, the Expressive and Improvisation conditions exhibited high agreement, suggesting that emotional intent was more effectively conveyed in these contexts. These results indicate that when performers play expressively and with more freedom, it becomes significantly easier to establish an emotional connection with the listeners.

\begin{table}[!t]
\centering
\begin{tabular}{c c c}
\hline
\underline{Condition} & \underline{P-value} & \underline{Alignment} \\ 
Modes & \textbf{0.0003} & Disagreement \\ 
Improvisation & 0.7616 & Agreement \\ 
\hline
Non-Expressive & \textbf{0.0136} & Disagreement \\ 
Expressive & 0.8086 & Agreement \\ 
\hline
\end{tabular}
\caption{Comparison of the performers' explicit emotions with the listeners' responses under specific conditions.}
\label{tab:chi_square_results}
\end{table}

\subsubsection{Implicit Emotional Analysis}

\begin{figure}[!b]
\centerline{\includegraphics[width=0.48\textwidth]{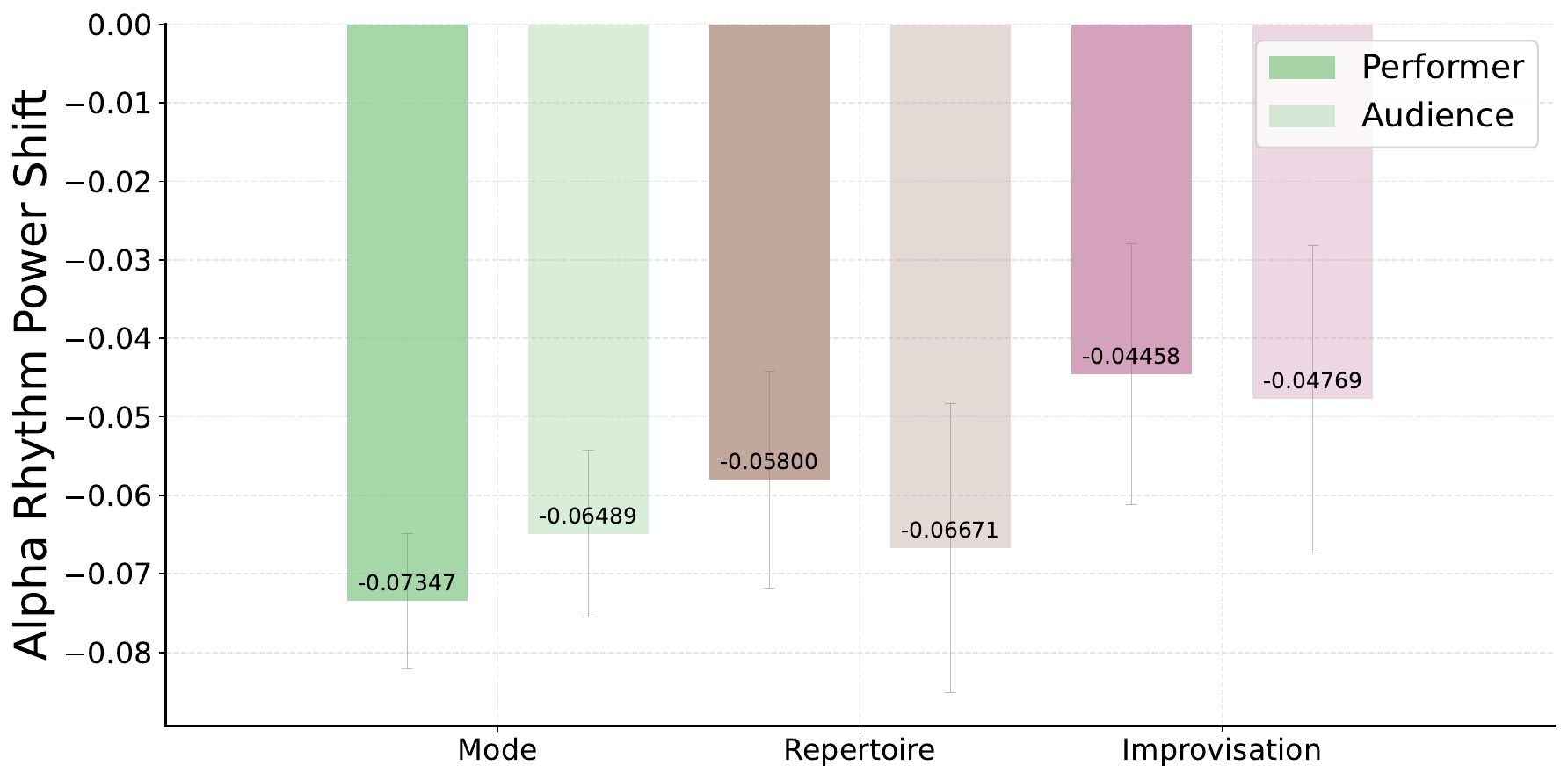}}
\caption{Alpha rhythm power analysis across performance conditions.}
\label{fig:eeg_conditions}
\end{figure}

The bar graph in Figure~\ref{fig:eeg_conditions} shows the comparison of alpha rhythm power changes between performer and audience across different conditions. In the Mode condition, there's the largest difference between performer and audience. The Repertoire condition shows similar values between performer and audience, with the audience showing a slightly stronger decrease. The Improvisation condition demonstrates the smallest overall decrease in alpha rhythm for both performer and audience, with values being closest to zero. Interestingly, the pattern shows a gradual reduction in the magnitude of change from Mode to Improvisation, suggesting that different musical conditions might require varying levels of neural engagement. When examining expressive and non-expressive performance conditions, both performer and audience showed similar alpha rhythm decreases. The error bars indicate overlap in the measurements across all conditions, suggesting that the observed differences were not statistically significant.

\subsection{Machine Learning Analysis }\label{subsec:emotion_transmission}

\begin{figure*}[!h]
    \centering
    \begin{minipage}{0.46\textwidth}
        \centering
        \includegraphics[width=\textwidth]{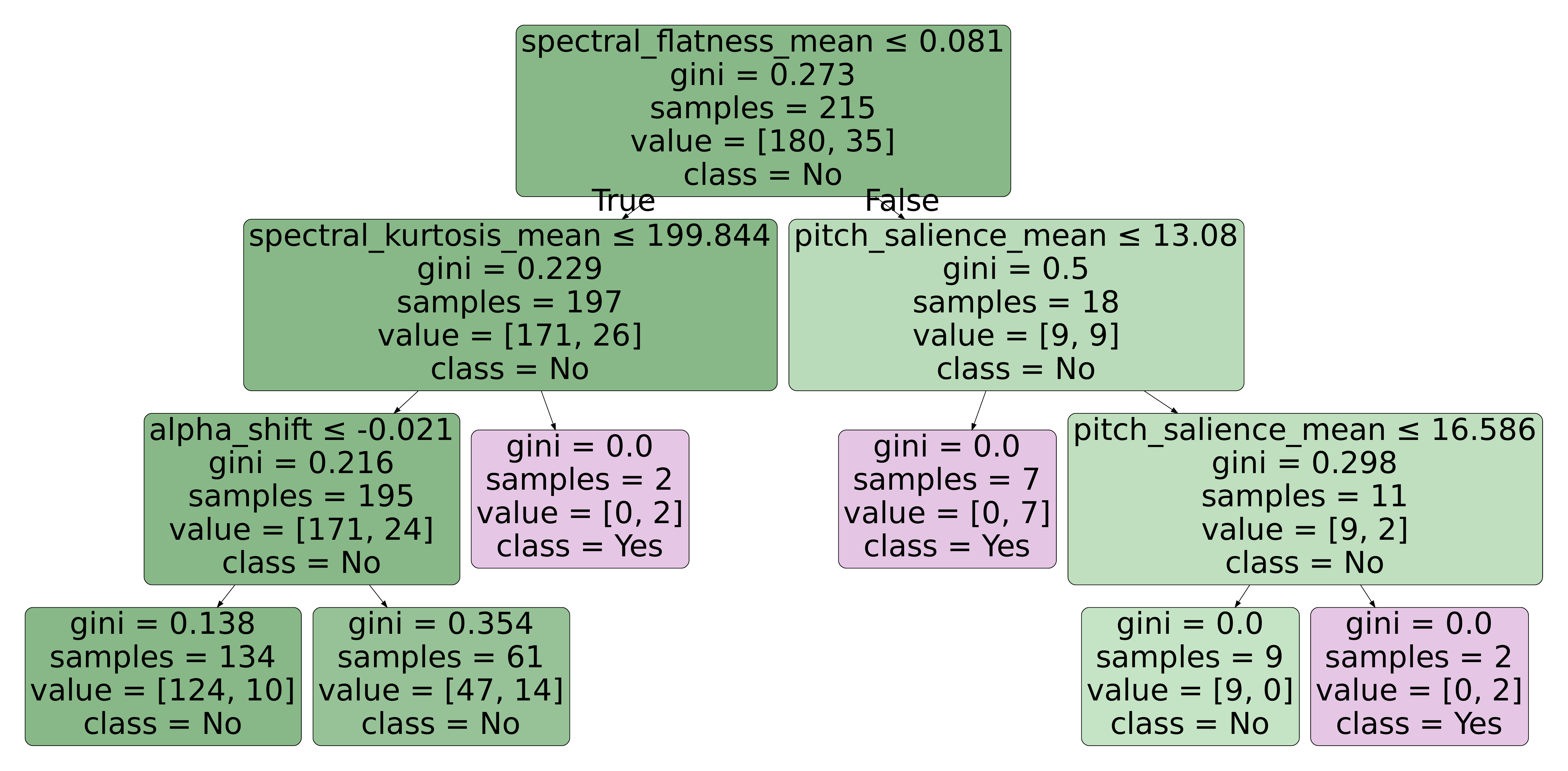}
        \label{fig:performer_peacefulness}
    \end{minipage}
    \hfill
    \begin{minipage}{0.50\textwidth}
        \centering
        \includegraphics[width=\textwidth]{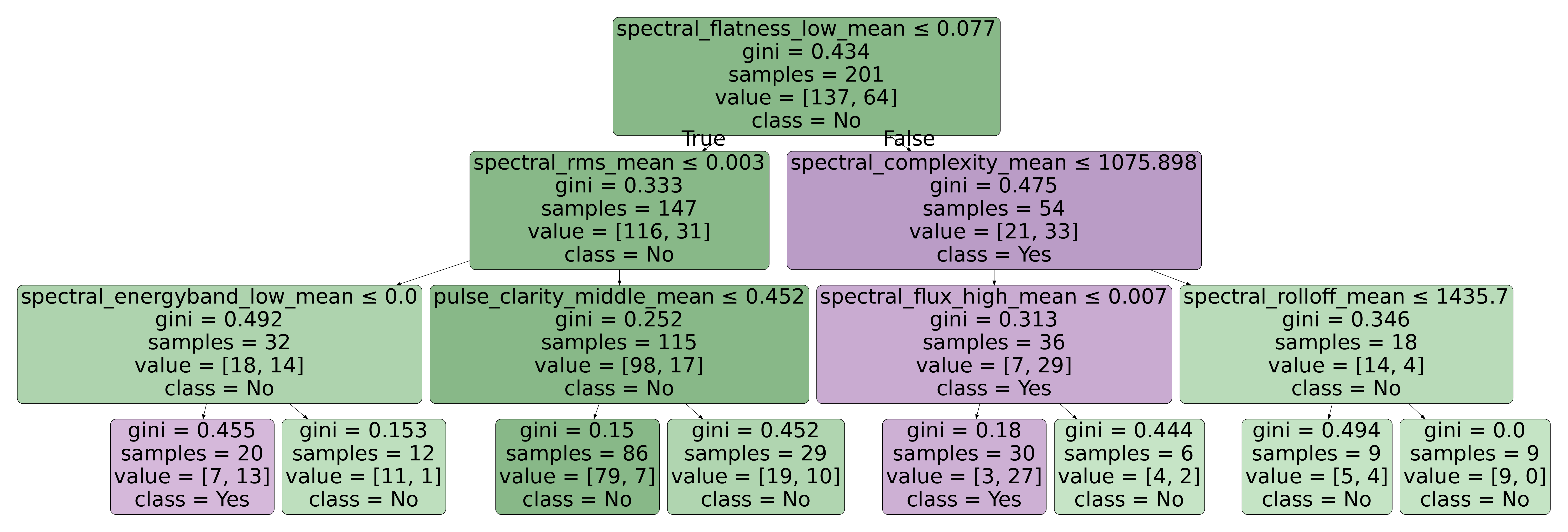}
        \label{fig:audience_peacefulness}
    \end{minipage}
    \caption{\textsc{Decision Trees} for \textit{Peacefulness} emotion tag: Performer (left) and Audience (right).}
    \label{fig:peacefulness_comparison}
\end{figure*}

The \textsc{Decision Trees} (see Figure~\ref{fig:peacefulness_comparison}) reveal distinct patterns in how performers and audiences perceive emotions in music, as illustrated through the example of \textit{Peacefulness}. The performer's tree (left) primarily relies on spectral features and incorporates neurophysiological data (alpha shift), suggesting their perception of emotional communication is influenced by both tonal characteristics and their physiological state. In contrast, the audience's tree (right) exhibits greater complexity with additional features like pulse clarity and spectral complexity, indicating that a broader range of acoustic factors influence their emotional perception. These structural differences highlight the unique perspectives and processing mechanisms between performers and listeners in musical experiences.

The \textsc{Decision Trees} models demonstrated robust predictive capabilities across all emotion tags. The performer model achieved an overall accuracy of 89.8\%, while the audience model reached 85.9\% accuracy. These high performance metrics suggest that both trees effectively capture the relationship between acoustic features, physiological data and emotional perception, with the performer model showing slightly better predictive power. 

\section{Discussion}\label{sec:discussion}

We constructed a novel experimental paradigm incorporating music performance and music listening of different types of musical structures, in order to study the musical and neural indices of emotion expression and perception in a quasi-naturalistic setting. Professional musicians were required to perform music based on a pre-selected repertoire, diatonic modes, as well as free improvisation. Both the musicians themselves and non-musician audience rated the emotions expressed in the music. 

With regard to our first research question, expressive and free performances exhibited sharper attacks, more notes, and greater spectral variations compared to non-expressive performances. This is in line with previous work. For instance, McAdams et al.~\cite{mcadams2017perception} found that higher tension arousal in music is associated with brighter sounds, increased spectral variation, and gentler attacks, while greater energy arousal is linked to brighter sounds with higher spectral centroids and slower spectral slope decreases. These findings could be explained by the strong emotional reactions that spectral variations evoke. In particular,~\cite{bannister2020vigilance} found that increasing loudness in an expressive musical piece significantly increased the frequency of chills that the listener experience, suggesting that dynamic intensity plays a key role in emotional engagement. This highlights the complex interaction between timbral characteristics and emotional responses. Finally, our results showed that when performers played expressively or improvised, the emotional connection with the audience was the highest. 

Following our second research question, our explicit emotional analysis demonstrated that expressive and improvisational performances elicited the strongest emotional reactions from the audience, with high-arousal emotions such as \textit{Power} and \textit{Tension} being more prevalent in these conditions. Moreover, Chi-Square tests confirmed that performer-intended emotions aligned more closely with audience perceptions in expressive and improvisational performances than in non-expressive conditions. These findings suggest that expressiveness enhances the clarity and transmission of emotional content in music.

Concerning our third research question, on the neural level, there was a trend for higher alpha power in the improvisation setting compared to repertoire and mode. Brain oscillatory activity in the alpha frequency band has been shown to be critical for creativity ~\cite{benedek2014alpha, zioga2024role}. Increased alpha power has been previously associated with higher creativity in EEG experiments investigating the functional role of brain oscillations in the creative process ~\cite{zioga2024role, lopata2017creativity}. Alpha power has been thought to support inhibition of common ideas, thus helping to reach more uncommon ideas ~\cite{benedek2014alpha}. For instance, higher alpha power was found in musicians during musical improvisation, reflecting enhanced inhibition processes, and this effect was positively correlated with the quality of the improvisations ~\cite{lopata2017creativity}. Additionally, musical improvisation was found to be associated with decreased activity in lateral prefrontal regions, interpreted as reduced activation ~\cite{limb2008neural}. Collectively, these results suggest that creativity is linked to the suppression of common ideas, that ``reside'' in frontal regions, via the mechanism of enhanced alpha oscillations.

By using \textsc{Decision Trees}, we achieved high accuracy in predicting emotional tags for both performers and audiences, and the model’s interpretability also allowed us to identify important features. Specifically, low spectral flatness and low spectral flux were key in predicting the \textit{Peacefulness} tag. These features indicate stable, tonal sounds with minimal frequency variation, which align with the perception of peacefulness. For the audience, this suggests that more stable, tonal music is perceived as peaceful, while for performers, low spectral flatness similarly correlates with peacefulness. 

This work underscores the importance of incorporating multimodal approaches to understanding musical communication. By combining computational audio analysis, real-time annotation, and neurophysiological measures, we provide a comprehensive framework for investigating the interplay between music interpretation, emotional expression, and perception. 

\section{Conclusion}

This study demonstrates that musical expressivity plays a critical role in shaping audience perception and emotional engagement. Our findings reveal that expressive and improvisational performances exhibit distinct acoustic characteristics, evoking stronger emotional responses, and lead to greater performer-audience neural synchronization. These results suggest that expressivity facilitates deeper audience engagement.

Furthermore, our methodological approach - integrating emotion annotation, audio feature analysis, and biosignal recordings - provides a robust framework for studying musical expressivity. Through both statistical analyses and interpretable machine learning techniques, we examined the relationships between these different modalities. The observed alignment between explicit emotional reports and neural alignment patterns reinforces the validity of using multimodal techniques to assess emotional transmission in music.

Future work could expand upon this work by incorporating additional biosignal measures, exploring different musical genres, and investigating the musical background of the performers. Additionally, further analysis of musical features such as notes, key signatures, and harmonic structures, combined with advanced machine learning approaches, could provide deeper insights into the relationship between musical interpretation and emotional expression. This could include experimenting with different machine learning architectures and techniques to better model the complex interactions between performance features and emotional responses.

\section{Ethics Statement}

This study was conducted in accordance with the ethical principles outlined in the Declaration of Helsinki. All participants provided informed consent prior to their participation, and their anonymity was maintained throughout the study. Ethical approval was obtained from the local ethics committee of the National and Kapodistrian University of Athens. Participants were informed of their right to withdraw at any time, and all data were anonymized and securely stored to ensure confidentiality.
Additionally, our study adhered to best practices for responsible research conduct. The annotation and biosignal collection procedures were designed to minimize participant discomfort, and all collected data were used exclusively for research purposes. The professional musicians were fairly compensated for their participation, ensuring ethical and equitable treatment of contributors.

\begin{acknowledgments}
 The research project is implemented in the framework of H.F.R.I call “Basic research Financing (Horizontal support of all Sciences)” under the National Recovery and Resilience Plan “Greece 2.0” funded by the European Union –NextGenerationEU(H.F.R.I. Project Number: 15111 - Emotional Artificial Intelligence in Music Expression). We thank the audience participants for making this work possible.
\end{acknowledgments} 

\bibliography{smc2025bib}
	
\end{document}